\documentclass[11pt,a4paper]{article}
\pdfoutput=1
\usepackage{amssymb} \usepackage{amsmath} \usepackage{graphicx}
\usepackage{epsfig,latexsym}
\begin{document}

\def\bef{\begin{figure}}
\def\eef{\end{figure}}
\newcommand{\ans}{ansatz }
\newcommand{\be}[1]{\begin{equation}\label{#1}}
\newcommand{\beq}{\begin{equation}}
\newcommand{\ee}{\end{equation}}
\newcommand{\beqn}[1]{\begin{eqnarray}\label{#1}}
\newcommand{\eeqn}{\end{eqnarray}}
\newcommand{\bd}{\begin{displaymath}}
\newcommand{\ed}{\end{displaymath}}
\newcommand{\mat}[4]{\left(\begin{array}{cc}{#1}&{#2}\\{#3}&{#4}
\end{array}\right)}
\newcommand{\matr}[9]{\left(\begin{array}{ccc}{#1}&{#2}&{#3}\\
{#4}&{#5}&{#6}\\{#7}&{#8}&{#9}\end{array}\right)}
\newcommand{\matrr}[6]{\left(\begin{array}{cc}{#1}&{#2}\\
{#3}&{#4}\\{#5}&{#6}\end{array}\right)}
\newcommand{\cvb}[3]{#1^{#2}_{#3}}
\def\lsim{\raise0.3ex\hbox{$\;<$\kern-0.75em\raise-1.1ex
e\hbox{$\sim\;$}}}
\def\gsim{\raise0.3ex\hbox{$\;>$\kern-0.75em\raise-1.1ex
\hbox{$\sim\;$}}}
\def\abs#1{\left| #1\right|}
\def\simlt{\mathrel{\lower2.5pt\vbox{\lineskip=0pt\baselineskip=0pt
           \hbox{$<$}\hbox{$\sim$}}}}
\def\simgt{\mathrel{\lower2.5pt\vbox{\lineskip=0pt\baselineskip=0pt
           \hbox{$>$}\hbox{$\sim$}}}}
\def\unity{{\hbox{1\kern-.8mm l}}}
\newcommand{\eps}{\varepsilon}
\def\ep{\epsilon}
\def\ga{\gamma}
\def\Ga{\Gamma}
\def\om{\omega}
\def\omp{{\omega^\prime}}
\def\Om{\Omega}
\def\la{\lambda}
\def\La{\Lambda}
\def\al{\alpha}
\newcommand{\ov}{\overline}
\renewcommand{\to}{\rightarrow}
\renewcommand{\vec}[1]{\mathbf{#1}}
\newcommand{\vect}[1]{\mbox{\boldmath$#1$}}
\def\tm{{\widetilde{m}}}
\def\mcirc{{\stackrel{o}{m}}}
\newcommand{\Dm}{\Delta m}
\newcommand{\dm}{\varepsilon}
\newcommand{\tanb}{\tan\beta}
\newcommand{\nbar}{\tilde{n}}
\newcommand\PM[1]{\begin{pmatrix}#1\end{pmatrix}}
\newcommand{\up}{\uparrow}
\newcommand{\down}{\downarrow}
\def\omE{\omega_{\rm Ter}}
%
%%%%%%%%%%     mauri    %%%%%%%%%%%%%%%%%%%%%%%%%%%%%%%%%

\newcommand{\Dsusy}{{susy \hspace{-9.4pt} \slash}\;}
\newcommand{\DCP}{{CP \hspace{-7.4pt} \slash}\;}
\newcommand{\mc}{\mathcal}
\newcommand{\gr}{\mathbf}
\renewcommand{\to}{\rightarrow}
\newcommand{\gtc}{\mathfrak}
\newcommand{\wh}{\widehat}
\newcommand{\br}{\langle}
\newcommand{\kt}{\rangle}

%%%%%%%%%%%%%%%%%%%%%%%%%%%%%%%%%%%%%%%%%%%%%%%%%%%%%%%%%%

\def\lsim{\mathrel{\mathop  {\hbox{\lower0.5ex\hbox{$\sim$}
\kern-0.8em\lower-0.7ex\hbox{$<$}}}}}
\def\gsim{\mathrel{\mathop  {\hbox{\lower0.5ex\hbox{$\sim$}
\kern-0.8em\lower-0.7ex\hbox{$>$}}}}}
%%%%%%%%%%%%%%%%%%%%%%%%%%%%%%%%%%

\def\nn{\\  \nonumber}
\def\de{\partial}
\def\brf{{\mathbf f}}
\def\bbf{\bar{\bf f}}
\def\bF{{\bf F}}
\def\bbF{\bar{\bf F}}
\def\bA{{\mathbf A}}
\def\bB{{\mathbf B}}
\def\bG{{\mathbf G}}
\def\bI{{\mathbf I}}
\def\bM{{\mathbf M}}
\def\bY{{\mathbf Y}}
\def\bX{{\mathbf X}}
\def\bS{{\mathbf S}}
\def\bb{{\mathbf b}}
\def\bh{{\mathbf h}}
\def\bg{{\mathbf g}}
\def\bla{{\mathbf \la}}
\def\bmu{\mathbf m }
\def\by{{\mathbf y}}
\def\bmu{\mbox{\boldmath $\mu$} }
\def\bsig{\mbox{\boldmath $\sigma$} }
\def\bunity{{\mathbf 1}}
\def\cA{{\cal A}}
\def\cB{{\cal B}}
\def\cC{{\cal C}}
\def\cD{{\cal D}}
\def\cF{{\cal F}}
\def\cG{{\cal G}}
\def\cH{{\cal H}}
\def\cI{{\cal I}}
\def\cL{{\cal L}}
\def\cN{{\cal N}}
\def\cM{{\cal M}}
\def\cO{{\cal O}}
\def\cR{{\cal R}}
\def\cS{{\cal S}}
\def\cT{{\cal T}}
\def\eV{{\rm eV}}

\large
 \begin{center}
 {\Large \bf  `Exotic vector-like pair' of color-triplet scalars} \

 \end{center}

 \vspace{0.1cm}

 \vspace{0.1cm}
 \begin{center}
{\large Andrea Addazi}
\footnote{E-mail: \,  andrea.addazi@infn.lngs.it}
\end{center}

{\it \it Dipartimento di Fisica,
 Universit\`a di L'Aquila, 67010 Coppito AQ and
LNGS, Laboratori Nazionali del Gran Sasso, 67010 Assergi AQ, Italy}

\vspace{1cm}
\begin{abstract}
\large

We propose a minimal extension of Standard Model, 
generating a Majorana mass for neutron,
connected with a 
mechanism of 
Post-Sphaleron Baryogenesis.
We consider  an `exotic vector-like pair' of color-triplet scalars,
an extra Majorana fermion $\psi$, and a scalar field $\phi$, giving mass to  $\psi$.  
The vector-like pair is defined `exotic'  because of  a peculiar mass term
of the color-triplet scalars, violating Baryon number as $\Delta B=1$.
Such a mass term could be generated by exotic instantons in a class of
string-inspired completions of the Standard Model:
 open (un-)oriented strings attached between D-brane stacks
 and Euclidean D-branes. 
A Post-Sphaleron Baryogenesis is realized through $\phi$-decays into six quarks (antiquarks), 
or through $\psi$-decays into three quarks (antiquarks).
This model suggests some intriguing B-violating signatures, testable in the next future,
in
Neutron-Antineutron physics and LHC.
We also discuss limits from FCNC. 
  Sterile fermion can also be light as $1-100\, \rm GeV$. 
In this case, the sterile fermion could be (meta)-stable
and $n-\bar{n}$ oscillation can be indirectly generated by two 
$n-\psi$, $\psi-\bar{n}$ oscillations, without needing of an effective Majorana mass for neutron. 
Majorana fermion $\psi$ can be a good candidate for 
 WIMP-like dark matter.

\end{abstract}

\baselineskip = 20pt

\section{Introduction}

Has the neutron a Majorana mass or not?
This is not just an academic question.
Majorana himself proposed in  `37', that neutron could 
have a Majorana mass term $\delta m\, nn+h.c$ \cite{Majorana}. 
We do not know if Majorana understood
immediately the depth of his proposal;
but today we get that existence of a "Majorana's fermion" is related to 
baryon or lepton numbers' violations.
In particular, a Majorana mass for neutron implies 
a neutron-antineutron transition, violating baryon number 
by $\Delta B=2$ \cite{2,3,4}. 
The current limit on $n-\bar{n}$ 
 is $\tau_{n\bar{n}}=1/\delta m>0.86\times 10^{8}\, \rm s$
with $90\%$ C.L., implying $\delta m<7.7 \times 10^{-24}\, \rm eV$ \cite{NNbar}. 
This corresponds to a constraint $\mathcal{M}>~300\, \rm TeV$ on the effective operator 
$(udd)^{2}/\mathcal{M}^{5}$. 
This limit is particularly loose with respect to other rare processes 
violating Baryon or Lepton numbers:
$\tau_{n\bar{n}}>3\, yr$ for neutron-antineutron 
can be compared with 
$\tau_{p-decay}\sim 10^{34\div 35}\, \rm yr$ for the Proton decays, 
$\tau_{0\nu\beta\beta}>10^{25}\, \rm yr$ for neutrinoless double beta decays \cite{PDGB}. 
For these reasons, neutron-antineutron is becoming more and more 
an interesting challenge for model building \cite{Mohapatra3,Mohapatra2,Mohapatra1,Patra,Wise,Dvali,Cheung} \footnote{See also \cite{NN4} for a short discussion about Neutron-Antineutron physics as a test
of a new fifth force interaction (a more complete version is in preparation \cite{NN5}). }, 
also considering possibility in the next future to 
enhance best limit of a factor $100$: $\tau_{n\bar{n}}>10^{10}\, \rm s$,
testing $1000\, \rm TeV$ scale \cite{Phillips:2014fgb}.  
(For a recent review about phenomenology of Baryon and Lepton violations, see also \cite{Perez:2015rza}). 
In this paper, we would like to suggest a 
simple minimal model connecting 
the "Majorana's question" with 
a mechanism of  Baryogenesis.
Depending on the particular region
of the parameters, this model connects 
neutron-antineutron physics with 
LHC, 
predicting a new peculiar phenomenology
in collider physics. This model does not produce proton decay,
and FCNC can be sufficiently suppressed. 

The main model's feature: we introduce an `exotic' mixing mass term for a vector-like pair of color scalar triplets, 
violating baryon number as $\Delta B=1$, i.e one color-triplet scalar has a different 
baryon number with respect to the other triplet antiscalar by exactly one unit. 
One scalar triplet has $B=1/3$, 
and the other has $B=2/3$. 
We call this an `exotic vector-like pair'.
We propose that existence of a $\Delta B=2$ Majorana Mass could be
connected to a $\Delta B=1$ exotic mass term! 
In a broad sense, we have a see-saw mechanism for neutron, 
involving a non-diagonal mass matrix for scalars rather than fermions 
 \footnote{The see-saw mechanism type I for the neutrino was originally proposed by Minkowski  \cite{13}, M.Gell-Mann,
P.Ramond and R.Slansky \cite{14,15}, by Yanigida \cite{16}, R.Mohapatra and G.Senjanovic \cite{17}. Then, other mechanisms called  type II   \cite{seesaw2a,seesaw2b,seesaw2c,seesaw2d,seesaw2e} and type III \cite{seesaw3a, seesaw2b, seesaw2c}, have been proposed later.  } \footnote{Probably, the most similar mechanism of the 
one proposed here is in
\cite{NNp1,NNp2,NNp3,NN5}.
In this case Baryon number is violated 
by a baryonic 'RH neutron', with a B-violating Majorana mass term.}. 
This model is inspired by proposals
in \cite{Blu1,Ibanez1,Ibanez2,Addazi:2014ila,Addazi:2015rwa}: 
(NMS-)SM is obtained as a low energy limit of 
open (un)-oriented strings, attached between 
D-brane stacks and Euclidean D-branes.
Euclidean D-branes are 
 exotic stringy instantons, that
can induce new non-perturbative mass terms,
violating vector-like $U(1)$s, rather than axial-ones.
In particular, in \cite{Addazi:2014ila,Addazi:2015rwa}, $R$-parity is {\it dynamically} broken by exotic instantons, 
producing only particular B-violating operators, such as a mass term for a vector-like pair
Proton decay is automatically suppressed in this model \cite{Addazi:2014ila,Addazi:2015rwa}.

An exotic vector-like pairs could be not only indirectly searched in $n-\bar{n}$ physics,
but also at LHC,
with peculiar processes:
$pp\rightarrow jjE_{T}\kern-14pt\slash\,\,\,\,\,$, for example, could be a spectacular signature
of exotic vector-like pairs and dark matter. 

This model can connect Neutron-Antineutron oscillations to Dark Matter problem rather
than to Baryogenesis. Infact, if $\psi$ is a metastable fermion of mass $1-1000\, \rm TeV$,
an exchange of a virtual exotic vector-like pair  can generate $n-\psi$ and $\psi-\bar{n}$ oscillations, with $\tau_{n-\psi}\simeq \tau_{\psi-\bar{n}}\simeq \tau_{n-\bar{n}}/2\simeq 10^{8}\, \rm s$.
In this case a Neutron-Antineutron transition can be generated as a combination of these two $|\Delta B|=1$ oscillations,
without needing of a Majorana mass for Neutron. $\psi$ can be a good candidate of WIMP Dark Matter.

The paper is organized as follows:
in Section 2, we describe the model for a Majorana neutron
also discuss suppression of FCNCs; 
in Section 3, we discuss implications for LHC physics;
in Section 4, connections with Baryogenesis;
in Section 5, we discuss a possible string-inspired scenario
for the effective model proposed, 
in Section 6, we present our conclusions. 

\section{A Model for a Neutron Majorana mass}
\begin{figure}[t]
\centerline{ \includegraphics [height=6cm,width=0.8 \columnwidth]{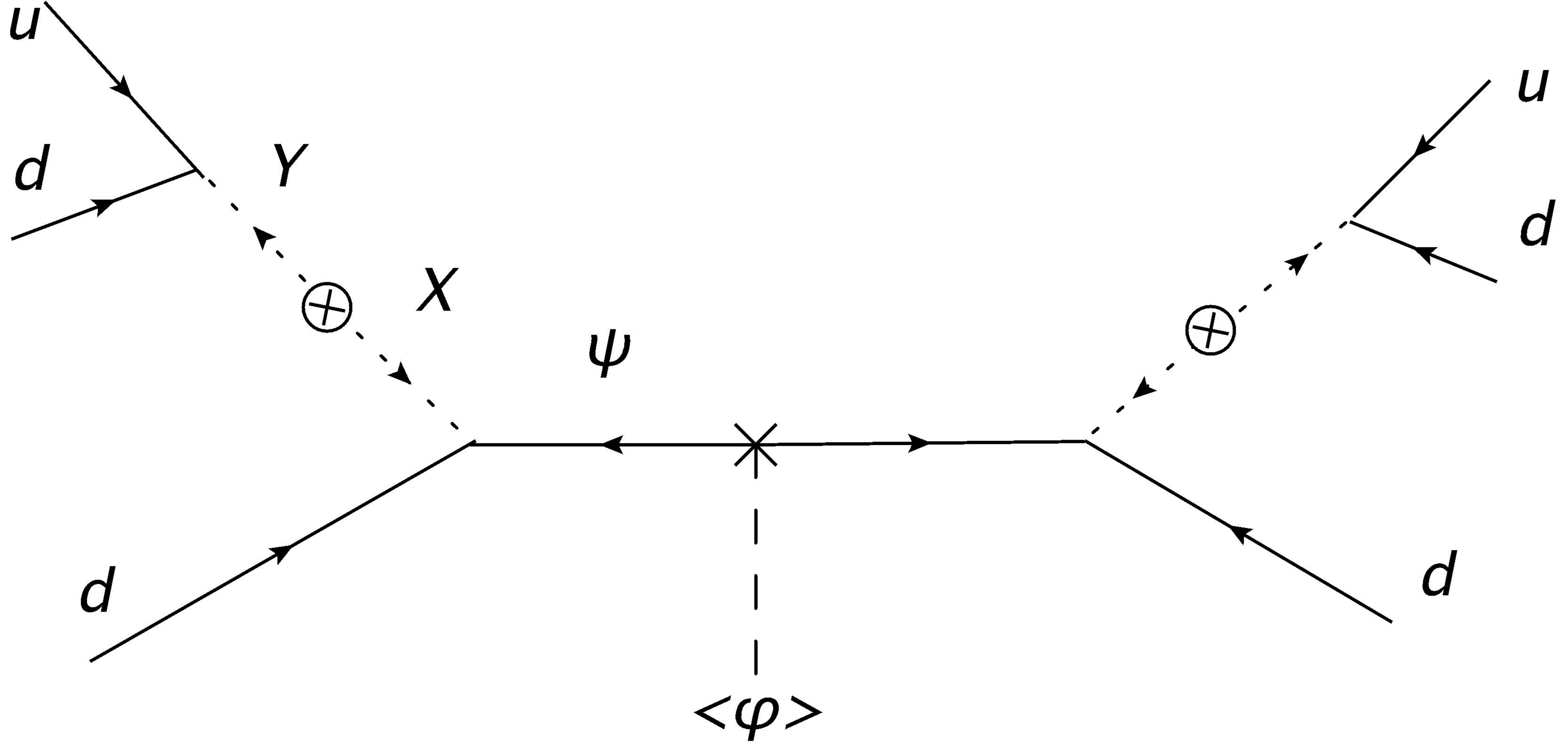}}
\vspace*{-1ex}
\caption{Diagram inducing a Neutron-Antineutron transition. The white blobs indicate the 
mixing mass term between the vector-like pair of color scalar triplets $\mathcal{X},\mathcal{Y}$. The central propagator is the Majorana fermion $\psi$. }
\label{plot}   % \ref{plot}
\end{figure}
We introduce a vector-like pair of (complex) color-triplet scalars $\mathcal{X}_{i},\mathcal{Y}^{i}$
(an their antiparticles)
with $i,j$ color indices of $SU(3)_{c}$. 
$\mathcal{X}$ has hypercharge $Y(\mathcal{X})=-2/3$, 
$\mathcal{Y}$ has hypercharge $Y(\mathcal{Y})=+2/3$.
Baryon and Lepton numbers are
$B(\mathcal{X})=1/3$, $B(\mathcal{Y})=2/3$ and 
$L(\mathcal{X})=L(\mathcal{Y})=0$. 
We also consider a Majorana sterile particle $\psi(1,1;0)$, 
with a mass term $\mu \psi \psi+h.c$. 
This is a gauge singlet with zero Baryon number, zero Lepton number,
zero hypercharge. 
\begin{table}[htbp]
\centering 
\begin{tabular}{||l|rl|l||}
\hline
\parbox{0.2\textwidth}{Fields}  & \parbox{0.1\textwidth}{$\,\,\,\,Y$} &  \parbox{0.1\textwidth}{$\,\,\,\,B$} & \parbox{0.1\textwidth}{$\,\,\,\,L$}      \\
\hline
\parbox{0.2\textwidth}{$\mathcal{X}(3,1;-2/3)$}   
    &  \parbox{0.1\textwidth}{$-2/3$}   &    \parbox{0.1\textwidth}{$+1/3$} &   \parbox{0.1\textwidth}{$\,\,\,\,0$}  \\
\parbox{0.2\textwidth}{$\mathcal{Y}(\bar{3},1;+2/3)$}  & \parbox{0.1\textwidth}{$+2/3$}   &    \parbox{0.1\textwidth}{$+2/3$} & \parbox{0.1\textwidth}{$\,\,\,\,0$}      \\
\parbox{0.2\textwidth}{$\psi(1,1;0)$}   & \parbox{0.1\textwidth}{$\,\,\,\,\,\,0$} & \parbox{0.1\textwidth}{$\,\,\,\,\,\,0$} & \parbox{0.1\textwidth}{$\,\,\,\,0$} \\
\parbox{0.2\textwidth}{$q_{L}(3,2;+1/3)$}   & \parbox{0.1\textwidth}{$+1/3$} & \parbox{0.1\textwidth}{$+1/3$} & \parbox{0.1\textwidth}{$\,\,\,\,0$} \\
\parbox{0.2\textwidth}{$u_{R}(\bar{3},1;-4/3)$}   & \parbox{0.1\textwidth}{$-4/3$} & \parbox{0.1\textwidth}{$-1/3$} & \parbox{0.1\textwidth}{$\,\,\,\,0$} \\
\parbox{0.2\textwidth}{$d_{R}(\bar{3},1;+2/3)$}   & \parbox{0.1\textwidth}{$+2/3$} & \parbox{0.1\textwidth}{$-1/3$} & \parbox{0.1\textwidth}{$\,\,\,\,0$} \\
\parbox{0.2\textwidth}{$l_{L}(1,2;-1)$}   & \parbox{0.1\textwidth}{$-1$} & \parbox{0.1\textwidth}{$\,\,\,\,\,\,0$} & \parbox{0.1\textwidth}{$-1$} \\
\parbox{0.2\textwidth}{$e_{R}(1,1;2)$}   & \parbox{0.1\textwidth}{$+2$} & \parbox{0.1\textwidth}{$\,\,\,\,\,\,0$} & \parbox{0.1\textwidth}{$+2$} \\
\hline
\end{tabular}
\caption{New matter fields introduced with respect to SM. We report their representation with respect to SM gauge group
$SU(3)\times SU(2) \times U(1)_{Y}$, 
their hypercharges $Y$ and their Baryon and Lepton numbers $B,L$. We also report Standard quarks and leptons for 
a comparison.  }
\end{table}

These fields, compatible with gauge invariances, can interact with quark fields as
\be{interactions}
\mathcal{L}_{Y}=y_{1}\mathcal{X}_{i}\psi d_{R}^{i} +y_{2}\mathcal{Y}^{i}u_{R}^{j}d_{R}^{k}\epsilon_{ijk}+h.c
\ee
mass terms for $\mathcal{X}$ and $\mathcal{Y}$,
\be{mass}
\mathcal{L}_{mass}=m_{\mathcal{X}}^{2}\mathcal{X}^{\dagger}\mathcal{X}+m_{\mathcal{Y}}^{2}\mathcal{Y}^{\dagger}\mathcal{Y}+h.c
\ee
and $\mathcal{X}-\mathcal{Y}$ has a peculiar mixing mass term 
\be{massmix}
\mathcal{L}_{\mathcal{X}-\mathcal{Y}}=\mathcal{M}_{0}^{2}\mathcal{X}^{i}\mathcal{Y}_{i}+h.c=\frac{1}{2}\mathcal{M}_{0}^{2}\epsilon_{ijk}\mathcal{X}^{i}\mathcal{Y}^{[jk]}+h.c
\ee
With these interactions, one can construct a Neutron-Antineutron transitions 
as shown in Fig.1. Note that all interactions terms are B-preserving, exception 
for mixing term $\mathcal{M}_{0}^{2}\epsilon_{ijk}\mathcal{X}^{i}\mathcal{Y}^{[jk]}$, violating baryon number as $\Delta B=1$.
Effective operator $(udd)^{2}/\mathcal{M}^{5}$ has a mass scale
$\mathcal{M}=(\mathcal{M}_{0}^{4}\mu)^{1/5}$, times coupling constants $y_{1,2}$, 
where $\mu$ is mass of fermion $\psi$. 
Experimental bound on $n-\bar{n}$ implies $\mathcal{M}>300\, \rm TeV$.
So, one can consider different choices of parameters $\mathcal{M}_{0}$
and $\mu$ in order to satisfy experimental limits. 
A trivial choice could be $\mathcal{M}_{0}=\mu=300\, \rm TeV$,
automatically saturating the bound. 
On the other hand, we can also consider for example 
$\mathcal{M}_{0}\simeq 1-10\, \rm TeV$
and $\mu \simeq 10^{6\div 10}\, \rm TeV$,
generating a lot of interesting physics for LHC, 
as discussed later. 
Another branch could be $\mu \simeq 1-10^{3}\, \rm GeV$
corresponding to $\mathcal{M}_{0}\simeq 7\times 10^{3\div 2}\rm TeV$. 
In this last case, the fermion $\psi$ is a natural candidate 
for WIMP dark matter, and Feynman diagram 
in Fig.1 can be seen as a combination of two oscillations $n-\psi$ and $\psi-\bar{n}$
with $\tau_{n\psi}\simeq \tau_{\psi\bar{n}}\simeq 10^{8}\, \rm s$:
$\psi$ is a (meta)stable particle, and not a virtual one in
propagator, in this case. Note that actual best limits 
on $n-\psi$ oscillations are $\tau \geq 414\, \rm s$, from Ultra Cold Neutron experiments, 
in condition of suppressed magnetic fields $|\mathcal{B}|<10^{-4}\, \rm Gauss$ \cite{Serembrov}. 

More precisely, in estimation of $\mathcal{M}$, we have to consider
not $\mathcal{M}_{0}^{2}$, but the smallest mass eigenvalue of mass matrix
of $\mathcal{X},\mathcal{Y}$. We assume $\mathcal{M}_{0}$ 
as a real parameter. 
We can decompose the color complex scalars as
 $\mathcal{X}=\frac{1}{\sqrt{2}}(\mathcal{X}_{1}+i\mathcal{X}_{2})$
and $\mathcal{Y}=\frac{1}{\sqrt{2}}(\mathcal{Y}_{1}+i\mathcal{Y}_{2})$,
and we can write mass matrix, in basis $(\mathcal{X}_{1},\mathcal{X}_{2},\mathcal{Y}_{1},\mathcal{Y}_{2})$
as 
\be{MassMat}
 M^{2}_{\rm eff} = \left( \begin{array}{cccc} m_{\mathcal{X}}^{2}& 0 & \mathcal{M}_{0}^{2}& 0
\ \\ 0 & m_{\mathcal{X}}^{2} & 0 & -\mathcal{M}_{0}^{2} \ \\
\mathcal{M}_{0}^{2} & 0 & m_{\mathcal{Y}}^{2} & 0 \ \\
0 & -\mathcal{M}_{0}^{2} & 0 & m_{\mathcal{Y}}^{2}
\end{array} \right)
\ee 
The eigenvalues are
\be{e12}
 \lambda_{\pm}^{2}=\frac{1}{2}\left(m_{\mathcal{X}}^{2}+m_{\mathcal{Y}}^{2}\pm \sqrt{4\mathcal{M}_{0}^{4}+(m_{\mathcal{X}}^{2}-m_{\mathcal{Y}}^{2})^{2}}\right)
\ee
(two-two degeneracies, as manifest in (\ref{MassMat})). 

In this model, we are not generating a proton decay process, if the mass of $\psi$ is higher than proton mass 
\footnote{We assume that other possible interactions of $\mathcal{X},\mathcal{Y},\psi$ with leptonic sector
are suppressed, in order to avoid other dangerous effective operators. 
For example, possible extra operators like $\mathcal{Y}q_{\alpha}l^{\alpha}$, 
leading to a proton decay operator $qqql/\Lambda^{2}$, can be avoided through opportune discrete symmetry 
$Z_{N}$, compatible with $\Delta B=1$ operators like $\mathcal{M}_{0}\mathcal{X}\mathcal{Y}$. 
Note that $\mathcal{X},\mathcal{Y}$ are not leptoquarks, they not have Lepton numbers, in our case. We are assuming that our model is not violating lepton number as
$\Delta L=1$; this is simple to realize just with a discrete symmetry $Z_{2}$. 
This can be also compatible with
  Majorana masses for neutrini $\Delta L=2$. 
  We also note that $\psi$ is not a Right-handed neutrino, 
  it has a Lepton number equal to zero. 
  For a complete classification of gauge discrete symmetries, 
  protecting the proton by $D=6$ operators,  for string constraints on Discrete symmetries, 
  see  \cite{discrete1,discrete2}.
  Alternatively, in a string-inspired model like \cite{Addazi:2014ila}, 
R-parity is dynamically broken by Exotic Instantons, generating (\ref{interactions})-(\ref{mass})-(\ref{massmix}), without other dangerous operators. 
For instance, $qqql/\Lambda^{2}$ is automatically avoided! \cite{Addazi:2014ila} }. 

\subsection{FCNC bounds and the space of the parameters}
 \begin{figure}[t]
\centerline{ \includegraphics [height=3.5cm,width=0.9 \columnwidth]{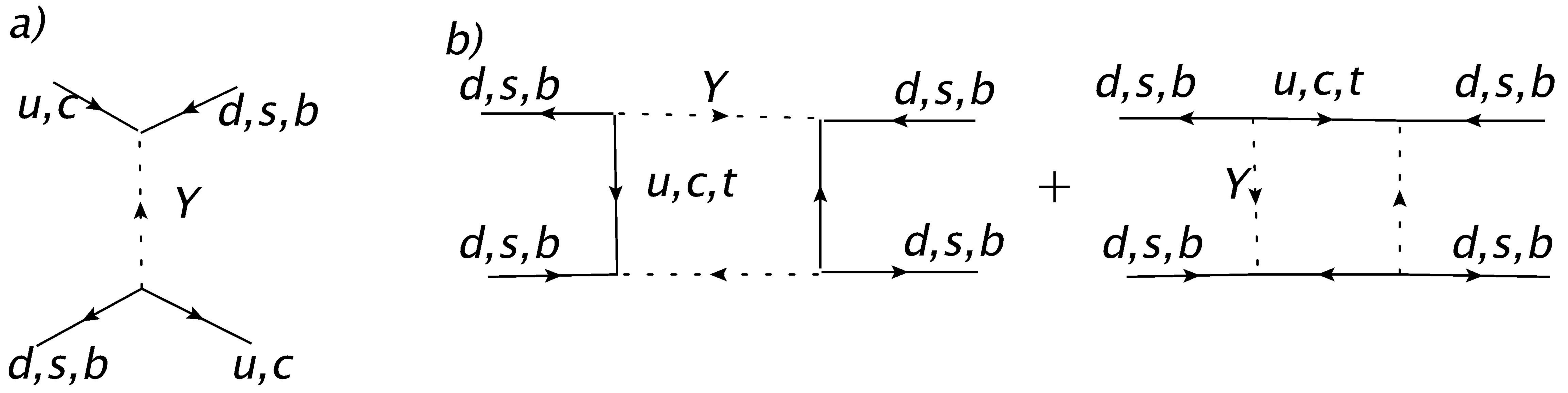}}
\vspace*{-1ex}
\caption{a) FCNCs tree-level diagrams mediated by $\mathcal{Y}$. 
b) Diagrams of neutral-meson oscillations, mediated by two $\mathcal{Y}$.   }  
\label{plot}   % \ref{plot}
\end{figure}
 \begin{figure}[t]
\centerline{ \includegraphics [height=3.5cm,width=0.9 \columnwidth]{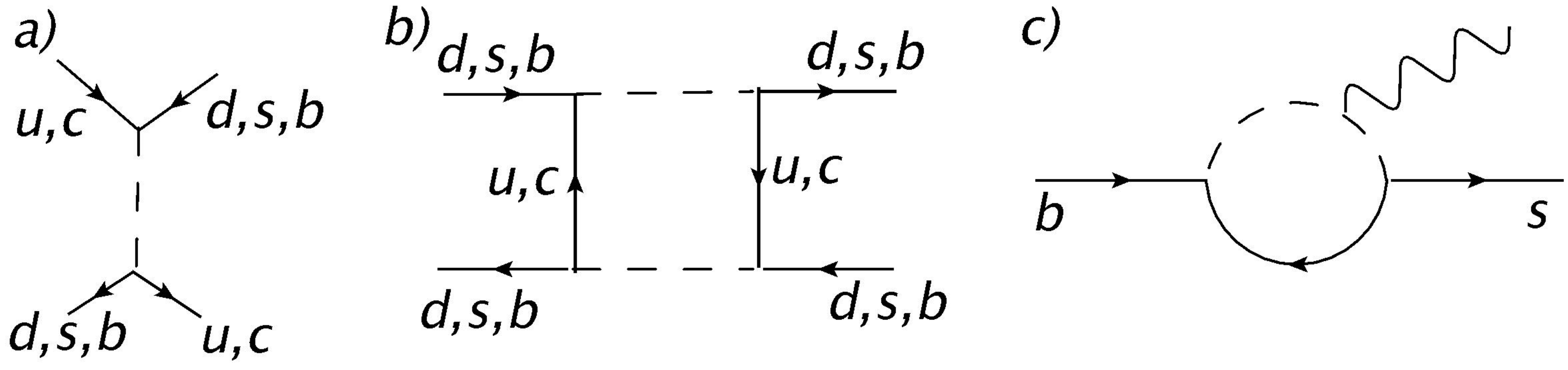}}
\vspace*{-1ex}
\caption{a) Diagram for meson decays into two mesons \cite{Addazi:2014ila}. This is mediated by two sterile fermions $\psi$ and four $\mathcal{X}-\mathcal{Y}$.
b) diagram for neutral meson-antimeson oscillation \cite{Addazi:2014ila}.  }  
\label{plot}   % \ref{plot}
\end{figure}
 \begin{figure}[t]
\centerline{ \includegraphics [height=3.5cm,width=0.8 \columnwidth]{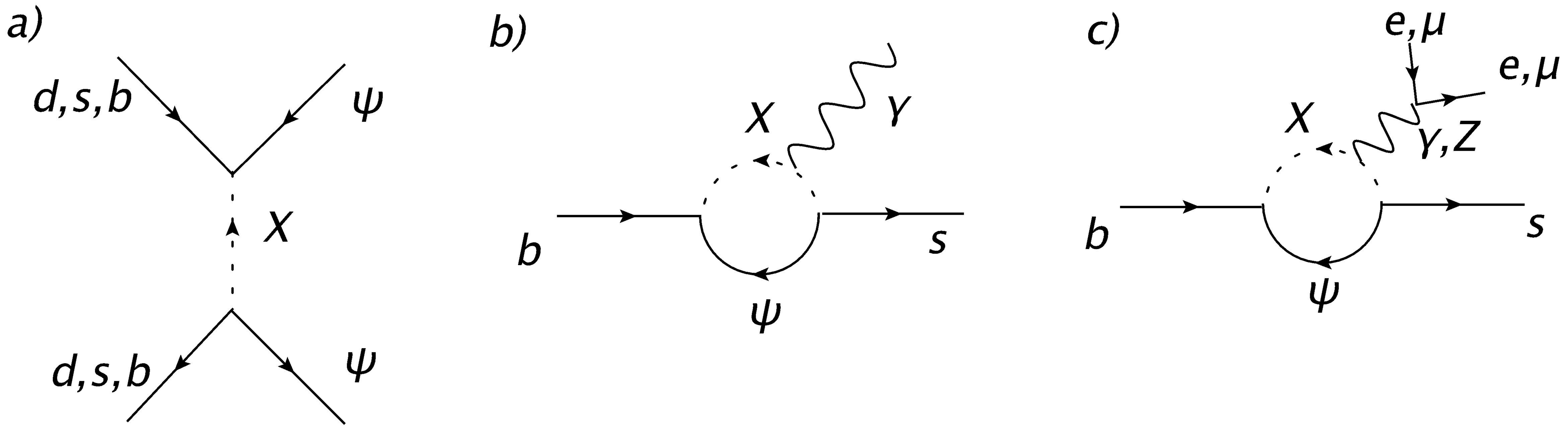}}
\vspace*{-1ex}
\caption{$B^{0}$,$B_{s}^{0}$,$\bar{B}^{0}$,$\bar{B}^{0}_{s}\rightarrow \psi\psi$ are possible if $2\mu\leq m_{B}$.
b) $b\rightarrow s\gamma$ transitions leading to $B \rightarrow K \gamma$, $\phi \gamma$. c) $b\rightarrow sl^{+}l^{-}$ transitions leading to $B \rightarrow K l^{+}l^{-},$ $\phi l^{+}l^{-}$.}  
\label{plot}   % \ref{plot}
\end{figure}
FCNC in meson physics are 
generated in our model. The strongest effects can come from a direct exchange of
one $\mathcal{Y}$, shown in Fig. 2. 
In particular diagrams (b) in Fig.2 contribute to neutral meson-antimeson oscillations 
such as  $K^{0}-\bar{K}^{0}$, $D^{0}-\bar{D}^{0}$, $B^{0}-\bar{B}^{0}$
etc. These constrain $\mathcal{Y}$'s mass up to $m_{\mathcal{Y}}\gtrsim 1000\, \rm TeV$.
However, these FCNC are not directly constraining $\mathcal{X}$'s mass.
In particular, assuming $m_{\mathcal{Y}}^{2}\simeq 10^{6}m_{\mathcal{X}}^{2}$
and $\mathcal{M}_{0}^{2}\simeq m_{\mathcal{X}}^{2}$, we obtain, from (\ref{e12}): 
$\lambda_{-}^{2}\simeq m_{\mathcal{X}}^{2}$ and $\lambda_{+}^{2}\simeq m_{\mathcal{Y}}^{2}$,
with mixing angles $\theta_{13}=\theta_{24}\sim 10^{-6}$.
So, mixings between $\mathcal{X}$ and $\mathcal{Y}$ are strongly suppressed 
in this case, but enough for neutron-antineutron transitions:
an prefactor of $10^{-12}$ in a $n-\bar{n}$ scale $(\mathcal{M}_{0}^{4}\mu)^{1/5}$.
has to be considered.
This strongly afflicts estimations of parameters:
for $\mathcal{M}_{0}=1-10\, \rm TeV$, it is enough a light $\psi$ of $\mu=1\div 100\, \rm GeV$!
As a consequence, the lightest eigenstate of mass matrix $(\ref{e12})$ can elude FCNC's constraints 
of Fig.2 and it can stay also near TeV scale.  Other FCNC's contributions,
directly involving $\mathcal{X}$, are suppressed, practically avoiding any current observations
as shown in Fig. 3. In neutral mesons' oscillations
$K^{0}-\bar{K}^{0}$, $D^{0}-\bar{D}^{0}$, $B^{0}-\bar{B}^{0}$
etc. any effects are suppressed as $\mathcal{M}_{0}^{-8}\mu^{-2}$. 
This strongly motivates a direct research of exotic color scalar triplets (the lightest eigenstate) at LHC. 
In next section, we will discuss these aspects. 
We also note that possible decays as $D^{0},B^{0}\rightarrow \psi\psi$, shown in Fig.4, can be generated 
if $2\mu \leq m_{D^{0},B^{0}}$. Suppose $2\mu\leq m_{B}$: 
in order to satisfy $n-\bar{n}$ limits, $\mathcal{X}$ may have $m_{\mathcal{X}}>> 1\, \rm TeV$,
strongly suppressing decays in Fig. 4, or colliders' processes. 
In the following discussion, we will assume $\mu>m_{B}/2\simeq 2.5\,\rm GeV$.

Other effects generated in our model are $b\rightarrow s\gamma$ and $b\rightarrow sl^{+}l^{-}$, shown in (b)-(c) Fig.4. Possible deviations in these are predicted in our model, with similar limits of supersymmetric models \cite{Kowalska:2014opa}, compatible with limits from the other channels discussed above. 

\section{LHC physics}
As discussed in Section 2, a direct production of the e.v.l.p is possible:
 bounds from neutron-antineutron physics allow $\mathcal{M}_{0}\sim 1-10\, \rm TeV$. A possible diagram of direct production 
of the lightest mass eigenstate 
of $\mathcal{X}-\mathcal{Y}$ is represented in Fig.5-(a). 
Compatible with FCNCs discussed above, 
We call two mass eigenstates as $\mathcal{Z}_{\pm}$, with mass eigenvalues $\lambda_{\pm}^{2}$. 
We can reach the lowest eigenstate $\mathcal{Z}_{-}$, with eigenvalue $\lambda_{-}\simeq \mathcal{M}_{0}$, compatible with FCNCs' bounds. For LHC physics, practically $\mathcal{Z}_{-}\simeq \mathcal{X}$. 
An interesting signature for LHC is $pp\rightarrow jjE_{T}\kern-12pt\slash\,\,\,\,$.
From this channel, we can put limit on $(m_{\mathcal{X}},m_{\psi})$;
essentially the same of  squarks $\tilde{q}\tilde{q}\rightarrow jjE_{T}\kern-14pt\slash\,\,\,\,\,$ \cite{ATLASL,CMSL}.
For $m_{\mathcal{X}}> 200\, \rm GeV$ $\rightarrow$ $\mu>200\, \rm GeV$;
$m_{\mathcal{X}}>500\, \rm GeV$ $\rightarrow$ $\mu>400\, \rm GeV$;
$m_{\mathcal{X}}>1000\, \rm GeV$ $\rightarrow$ $\mu$ is unbounded from below. 
As a consequence, $\psi$ could be a (meta)stable particle visible at LHC as transverse missing energy and Dark Matter Direct Detection.
 In this scenario, Neutron-Antineutron physics is directly connected to the Dark Matter question
\footnote{If $\psi$ compose all Dark Matter, from WIMP relic abundance $\mu>7\, \rm GeV$ \cite{WIMP}.   }. 

 \begin{figure}[t]
\centerline{ \includegraphics [height=6cm,width=1.0 \columnwidth]{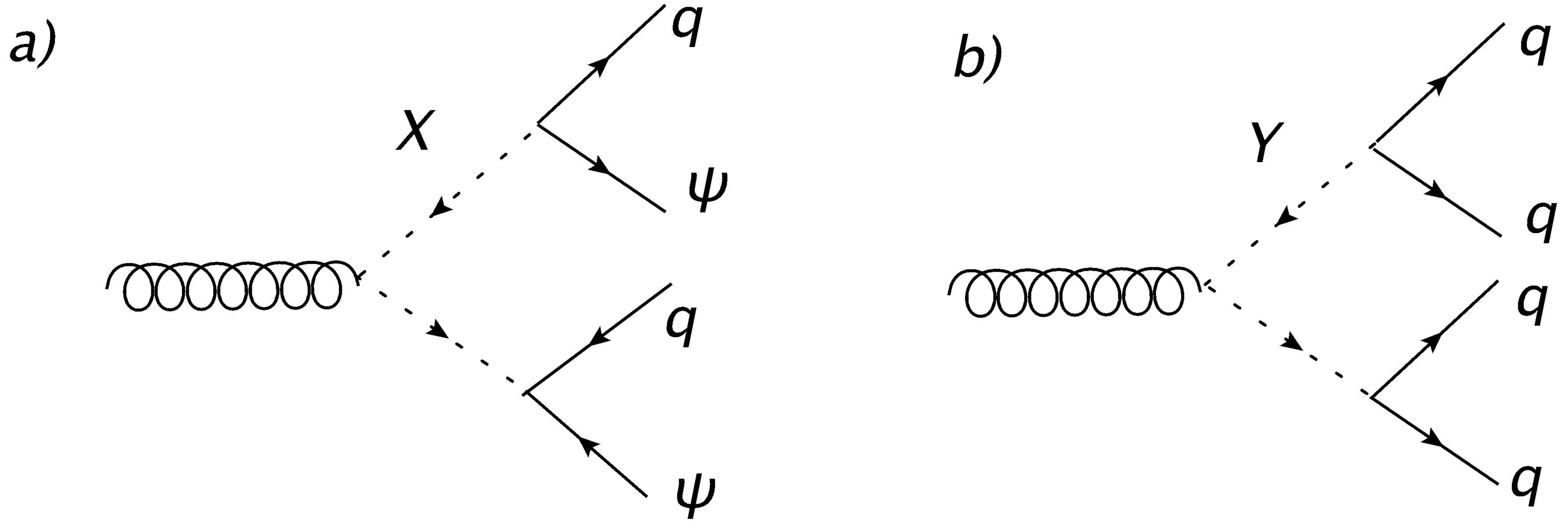}}
\vspace*{-1ex}
\caption{a) Missing energy channel $jjE_{T}\kern-14pt\slash\,\,\,\,\,$ at LHC; b) Diagram leading to $4j$ and $t\bar{t}jj$ channels.     }  
\label{plot}   % \ref{plot}
\end{figure}

We also mention limits from top-jet and di-jets channels, in Fig.3-(b),   around $1\, \rm TeV$ (top-jet $900\, \rm GeV$, di-jets $1.2\, \rm TeV$) \cite{33}, but these are not lower than FCNC ones cited above.

\section{Post-Sphaleron Baryogenesis}
In the proposed mode, one can envisage two simple mechanisms
for post-sphaleron baryogenesis: i) $\phi$-decays into six-quarks (antiquarks), ii) $\psi$-decays into three-quarks (antiquarks). 
We discuss these two in the following.

\subsection{Scalar-decays into six quarks (antiquarks)}

We can reverse diagram in Fig.1, considering 
the mass parameter of $\psi$ as generated by a scalar field $\phi$, acquiring a vev scale
$v$, with $\mu=y_{\psi}v$. 
For the moment, the mass of $\phi$ is a free parameter, $M_{\phi}$ \footnote{More precisely we can rewrite $\phi=(v+\phi_{r}+\phi_{i})/\sqrt{2}$, 
and the dynamical scalar decaying is $\phi_{r}$. In the following discussions, for $\phi$-decays we will always mean $\phi_{r}$-decays,
and for $M_{\phi}$ we will mean $M_{\phi_{r}}$. }. In Fig.6 we show decay diagrams, at tree level and one-loop. 
\begin{figure}[t]
\centerline{ \includegraphics [height=10cm,width=1.0 \columnwidth]{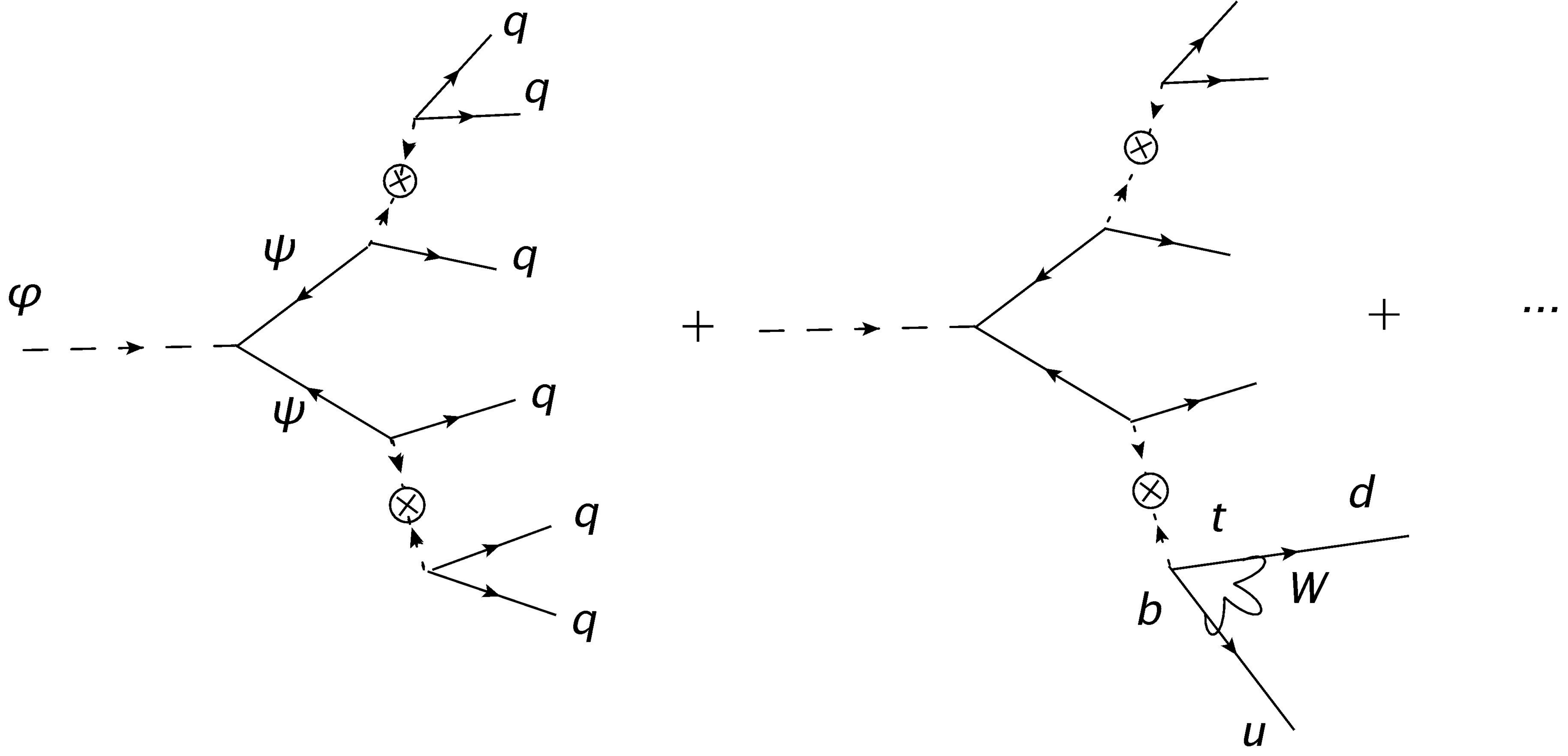}}
\vspace*{-1ex}
\caption{Decay $\phi \rightarrow 6q$: the first is a tree level contribution, but 
also one loops contributions, as the one shown, have to be considered.  One-loop contribution in figure is an example of electroweak CKM correction to decay amplitude
through an exchange of a $W$ boson, converting $top-down$ and $bottom-up$. 
Because of Majorana particle $\psi$, we can revert all arrows in Feynman diagrams,  obtaining 
$\phi \rightarrow 6\bar{q}$.   }
\label{plot}   % \ref{plot}
\end{figure}
We can evaluate the amplitude $\mathcal{M}$, at tree level, as
\be{tree}
\mathcal{M}^{tree}\simeq \frac{y_{\psi}<\phi>Tr[y^{\dagger}_{1}y_{1}]Tr[y^{\dagger}_{2}y_{2}]\mathcal{V}^{*}\mathcal{V}}{\mathcal{M}_{0}^{4}\mu^{2}}=\frac{Tr[y^{\dagger}_{1}y_{1}]Tr[y^{\dagger}_{2}y_{2}]\mathcal{V}^{*}\mathcal{V}}{\mathcal{M}_{0}^{4}\mu}
\ee
where $\mathcal{V}$ is the diagonalizing matrix of masses (\ref{MassMat}), suppressing 
the amplitude as $\mathcal{V}^{*}\mathcal{V}\sim 10^{-12}$, as cited above. 
under the assumption $\lambda_{+}>>\lambda_{-}\simeq \mathcal{M}_{0}$, where $\lambda_{\pm}$ are the mass eigenvalues in (\ref{e12}). 

One-loop corrections from the electroweak sector
can be evaluated as (assuming all the couplings in $\lambda_{1}\sim \lambda_{2}\sim 10^{-3}\div 1$)
\be{oneloop}
\mathcal{M}^{1-loop}\simeq c \mathcal{V}^{*}\mathcal{V} V_{ub}^{*}V_{td} \Phi_{1-loop}\bar{\mathcal{O}}^{2}_{n\bar{n}}\left(\frac{\mu m_{t}m_{b}}{m^{2}_{W}}\right)
\ee
where $c\simeq (10^{-3}\div 1)^4/128\pi^{2}$, and $\bar{\mathcal{O}}^{2}_{n\bar{n}} \equiv <\bar{n}|\mathcal{O}^{2}|n>\simeq -0.3 \times 10^{-5}\, \rm GeV^{6}$ from MIT bag model \cite{MIT} (confirmed also by recent lattice calculations \cite{Lattice}). 
$\Phi^{1-loop}$ (with dimension mass $[\Phi^{1-loop}]=M^{-4}$) is a function depending on the mass of the quarks closing the one-loop in Fig. 6
(the the top and bottom masses, in dominant contribution). However, there are also other possible contributions, closing 1-loops involving the vector-like pairs,
in which $\Phi^{1-loop}$ is depending also on vector-like pair.
A precise evaluation of such a formula is not necessary for our purposes. 
It is a good approximation to compare directly (\ref{tree}) with the present bounds on neutron-antineutron physics.
In principle, we have also to consider running prefactors connecting 
high energy physics of baryogenesis with low energy neutron-antineutron physics.
This prefactor is around $10^{-2}$ \cite{28}. 

At three level, the decay rate of $\phi$ is the square modulus of the amplitude (\ref{tree}), times a phase space factor 
for a $6q$ (or $6\bar{q}$) final state:
\be{Rate}
\Gamma_{\phi}=\Gamma(\phi \rightarrow 6q)+\Gamma(\phi \rightarrow 6\bar{q})=\mathcal{I}
\mathcal{V}^{*}\mathcal{V}Tr[y^{\dagger}_{1}y_{1}]^{2}Tr[y^{\dagger}_{2}y_{2}]^{2}\left(\frac{M_{\phi}^{13}}{\mu^{4}\mathcal{M}_{0}^{8}}\right)
\ee
with $\mathcal{I} \simeq 7\times 10^{-18}$ a numerical factor coming from a numerical integration in the phase space times combinatoric factors  (practically independent from the ratios of mass parameters, the variations
on this integration are of the order of $1\, \%$, not important for our purposes).

Considering the case of a Post-sphaleron baryogenesis: 
the rate (\ref{Rate}) has to be smaller than the Hubble rate 
at a temperature near the electroweak phase transition epoch:
$\Gamma_{S}<H(T_{ew})$. 
We consider a decay temperature indicatively between $100 GeV \div 200\, \rm MeV$, 
between electroweak phase transition and the QCD phase transition ($\Lambda_{QCD}\simeq 200\, \rm MeV$). The decay temperature $\bar{T}$ can be found solving 
the equation 
\be{HGamma}
\Gamma_{S}(\bar{T})\simeq H(\bar{T})\simeq 1.66 g_{*}^{1/2}\frac{\bar{T}^{2}}{M_{Pl}}
\ee
where $g_{*}$ is the number of degrees of freedom at $\bar{T}$.
From this we can get 
\be{Tbar}
\bar{T}\simeq \sqrt{\frac{M_{Pl}M_{\phi}^{13}}{(2\pi)^{9}\mu^{4}\mathcal{M}_{0}^{8}}}
\ee
So, a post sphaleron scenario impose limits on the masses' ratios.
For example, supposing $\bar{T}\sim 100 \div 200\, \rm GeV$ and 
$M_{\phi}\simeq   0.5\, \rm TeV$:
we can get bounds on the vector-like pair mixing mass $\mathcal{M}_{0}$ 
and Majorana fermion mass,
well compatible with the ones coming from neutron-antineutron physics. 

Finally, we can evaluate the primordial baryon asymmetry parameter, directly related to 
the observed baryon asymmetry:
\be{Asy}
\epsilon \simeq \frac{n_{\phi}}{n_{\gamma}}\frac{\Gamma(\phi \rightarrow 6q)-\Gamma(\phi \rightarrow 6\bar{q})}{\Gamma_{\phi}}
\ee
It is necessary to evaluate this including 1-loop CP-violating contributions 
coming from the electroweak sector, i.e CKM CP violating contributions. 
The contribution from 1-loop vertices \footnote{One can consider also 1-loop contributions coming involving also $\mathcal{X},\mathcal{Y},\psi$ in the propagators. However, one can numerically evaluates these contributions and discover that they are subdominant with respect to the contributions in (\ref{Asy}). Also Self-energy contributions (or wave-function renormalizations) give not important contributions for our estimations.} 
as the one shown in Fig. 6
are (considering (\ref{oneloop}) )
\be{epsilon}
\epsilon^{V}\simeq \frac{g_{2}^{2}}{32\pi } \frac{\mathcal{V}^{*}\mathcal{V} y_{2}^{\dagger}V_{td}^{*}V_{ub}  y_{2}}{ Tr[y_{2}^{\dagger} y_{2}]} \frac{m_{t}m_{b}}{m_{W}^{2}}\left[1+\frac{9m_{W}^{2}}{M_{\phi}^{2}}\rm ln\left(1+\frac{M_{\phi}^{2}}{3m_{W}^{2}} \right) \right]
\ee
With $\epsilon^{V}$ one-loop vertex contribution. 
So the asymmetry is controlled by $M_{\phi}$. As a consequence, 
$M_{\phi}>>500\div 1000\, \rm GeV$ suppresses the contribution from the vertex. 
Cimparing this bound with the other one coming from (\ref{Tbar}),
the region of the parameters discussed in Section 2 are well compatible.
As a consequence, a Post-sphaleron baryogenesis is possible 
and naturally predicts a neutron-antineutron oscillation 
of $\tau_{n\bar{n}}\simeq 10^{8}-10^{10}\, \rm s$. 
 
Finally, we also have to consider the dilution of the baryon asymmetry:
$\bar{T}\simeq M_{\phi}/5\div M_{\phi}/10$, 
the decay of $\phi$ generated entropy into the primordial plasma.
The dilution can be evaluate as the ratio of entropy density before and after $\phi$-decay:
\be{dilution}
\mathcal{D}=\frac{s_{initial}}{s_{final}}\simeq \frac{0.6\sqrt{\Gamma_{\phi}M_{Pl}}}{g_{*}^{1/4} M_{\phi} r_{\phi}}
\ee
where $r_{\phi}=n_{\phi}/s$ is at the decays' epoch. 
This can be estimated as
\be{roughly}
\mathcal{D}\sim k\frac{\bar{T}}{{M_{\phi}}}\sim k(10\%\div 20\%)
\ee
(where $k$ parametrize also extra suppressions from the couplings). 
From (\ref{epsilon}), we can find $\epsilon \sim 10^{-8\div 9}$,
but this has to be normalized with the dilution factor.
We obtain (assuming all couplings near one i.e $k\sim 1$), $\eta_{B}\sim \mathcal{D} \epsilon \sim 10^{-9\div 10}$,
where $\eta_{B}=(n_{b}-n_{\bar{b}})/n_{\gamma}$, 
as requeired observations ($\eta_{B}^{exp}=(6.04 \pm 0.08)\times 10^{-10}$
\cite{etaexp}). 

So, we can conclude that this mechanism can generate baryon asymmetry in our Universe,
during a Post-Sphaleron epoch, 
satisfying all Sakharov's conditions i.e i) out of thermal equilibrium; ii) CP-violating processes iii) B-violating processes
\cite{Sakarov}. 

\subsection{Majorana fermion decays in three quarks (antiquarks)}
Alternatively, we can consider directly $\psi\rightarrow u_{i}d_{j}d_{k},\bar{u}_{i}\bar{d}_{j}\bar{d}_{k}$, 
in which $\mu$ is below electroweak scale. In this scenario, color triplets cannot be detected at LHC. 
The decay rate can be evaluated as
\be{decayR2}
\Gamma_{\psi \rightarrow qqq,\bar{q}\bar{q}\bar{q}}=ck \mu^{5}\left(\frac{1}{\lambda_{+}^{2}}-\frac{1}{\lambda_{-}^{2}}\right)^{2}
\ee
where $\lambda_{\pm}$ are mass eigenvalues in (\ref{e12}),
and 
$$c=1/4096\pi^{3},\,\,\,\,k=\mathcal{V}^{*}\mathcal{V} y_{1}^{\dagger}y_{1}\rm Tr[y_{2}^{\dagger}y_{2}]$$ 
($c$ contains also color factor $6$ in numerator). 
We are assuming $\lambda_{\pm}>>\mu$. 
Under the assumption $\lambda_{+}>>\lambda_{-}\simeq \mathcal{M}_{0}$, 
(\ref{decayR2}) is simplified as 
\be{decayR2}
\Gamma_{\psi \rightarrow qqq,\bar{q}\bar{q}\bar{q}}=ck \mu^{5}\frac{1}{\mathcal{M}_{0}^{4}}
\ee
However, we have also to consider scattering processes.
$q+\psi \rightarrow \bar{q}\bar{q}$: they go-out of equilibrium at the same temperature $\bar{T}$ 
of $\psi \rightarrow 3q(\bar{q})$ decays. 
For $\bar{T}<\mu$, $\psi$ cannot be produced, for lack of phase space.
So, one has also to consider $\psi\psi \rightarrow q\bar{q}$ contributions
to baryon asymmetry generation. 
Extra one-loop electroweak corrections
($W^{\pm}$ exchanges) lead to dominant contributions as $(\ref{epsilon})$ cited above. 
From this, we can estimate $\epsilon \simeq 10^{-8}\div 10^{-9}$,
for $k\sim 1$ (natural couplings), ulteriorly suppressed by 
 by dilution factor for $10^{-1}$, as discussed in the previous subsection.
We conclude that also mechanism seems a viable way to 
generate the observed Baryon asymmetry. 

\section{Beyond the Toy-Model: String-Inspired Standard Model and Exotic Instantons}
In this section, we would like to discuss a possible explanation of the toy-model \footnote{We mention that, recently, a toy model for a supersymmetric non-local QFT
was discussed in \cite{Addazi:2015dxa}.}, 
as a String-Inspired class of model, embedding the Standard Model,
generating an exotic mass term for the vector-like pairs
 \begin{figure}[t]
\centerline{ \includegraphics [height=4cm,width=0.8 \columnwidth]{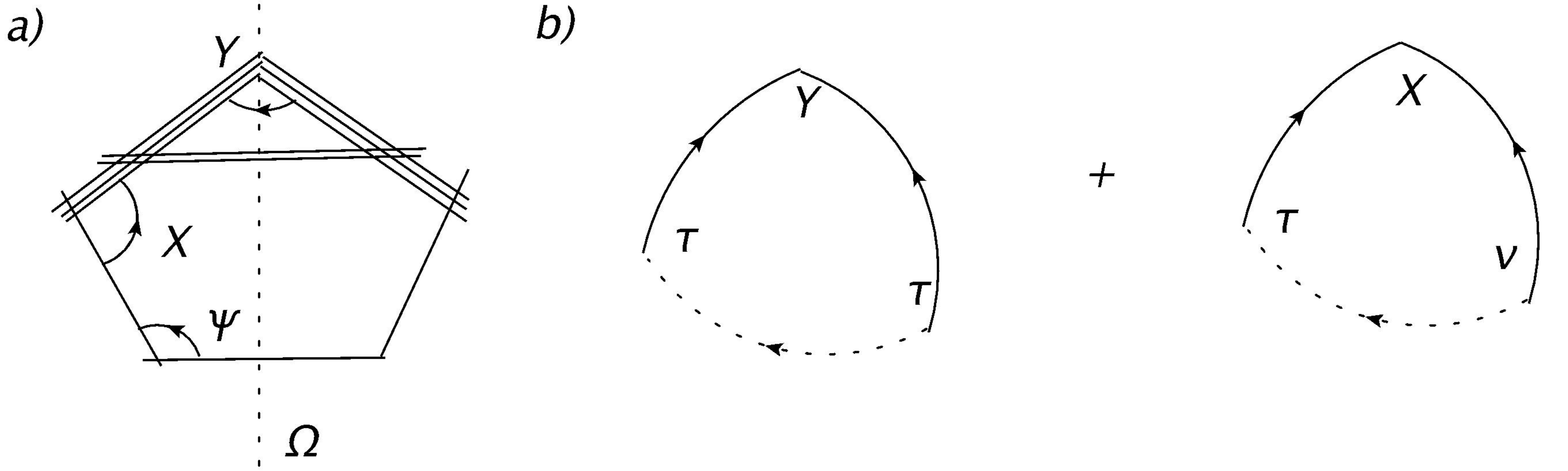}}
\vspace*{-1ex}
\caption{a)  (Sub)-system of D-branes stacks generating our toy-model content of fields at low energy limit. b) Mixed-Disk amplitudes generating an Exotic mass term for $\mathcal{X},\mathcal{Y}$.    }  
\label{plot}   % \ref{plot}
\end{figure}
 \begin{figure}[t]
\centerline{ \includegraphics [height=3.8cm,width=0.5 \columnwidth]{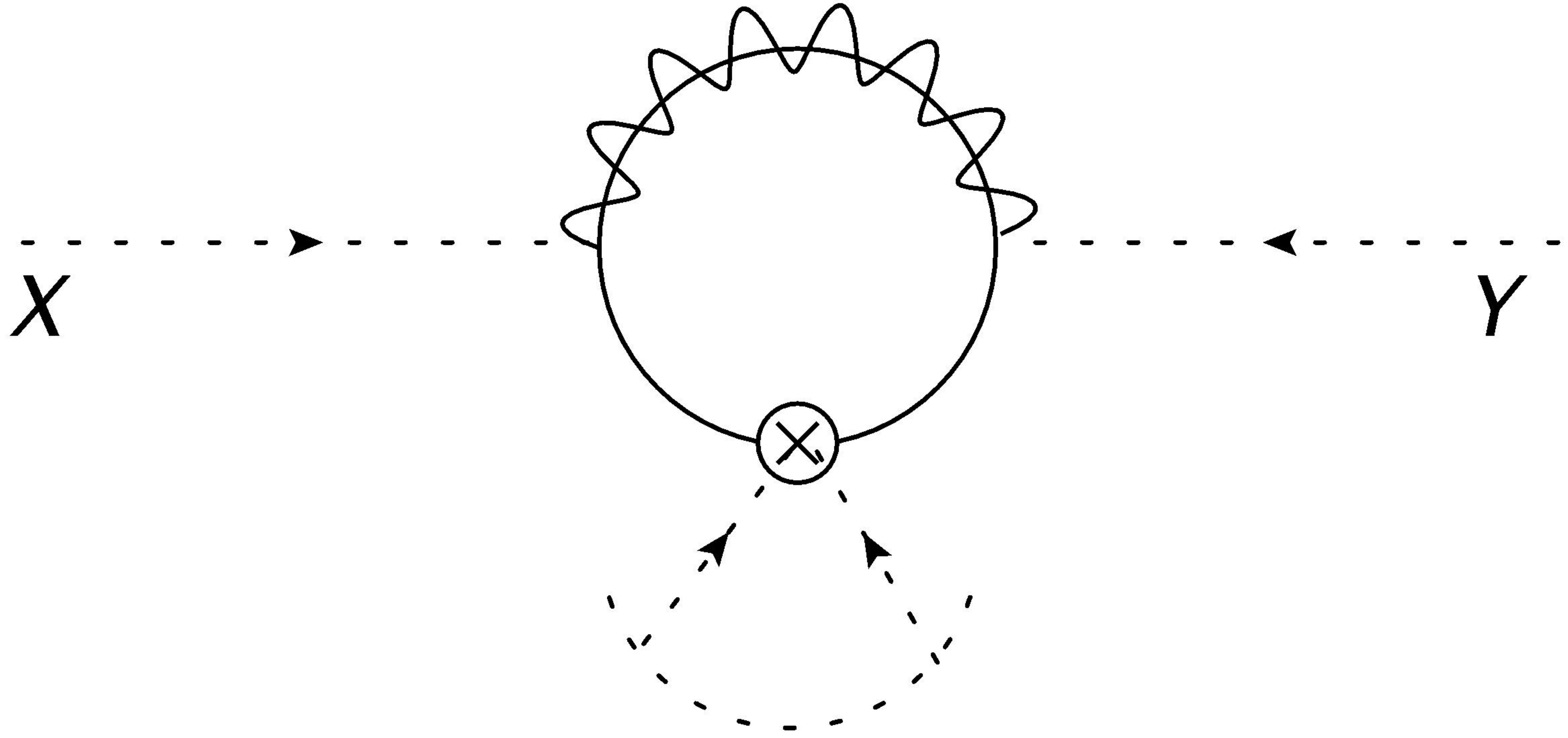}}
\vspace*{-1ex}
\caption{Exotic mass for $\mathcal{X},\mathcal{Y}$, generated by one-loop corrections,
containing one gaugino, and a $\psi_{\mathcal{X},\mathcal{Y}}$ mixing induced by Exotic Instantons 
(white blob with dashed lines).  }  
\label{plot}   % \ref{plot}
\end{figure}
We suggest a little different variant with respect to the one suggested in \cite{Addazi:2014ila}:
a IIA (un)-oriented string theory, with stacks of
D6 ordinary branes, and Euclidean  and D2-branes,
wrapping 3-cycles on $CY_{3}$, and an antisymmetric Mirror Plane $\Omega_{-}$,
recovering at low energy limit $U(3)\times U(2) \times U(1) \times U'(1)$
or $U(3)\times Sp(2) \times U(1) \times U'(1)$, $\mathcal{N}=1$ susy, R-parity preserving.
A possible simplified scheme of D-brane stacks (sub)-system is shown in Fig.7-(a):
$\mathcal{X},\mathcal{Y}$ are scalar parts of superfields ${\bf X},{\bf Y}$, 
attached between a $U(3)$ stack and a $U(1)$ stack, and a $U(3)$ stack and
 its mirror twin, with respect the mirror plane $\Omega$, respectively. 
On the other hand, $\psi$ is the fermionic part of a superfield ${\bf \Psi}$ 
living between two $U(1)$ stacks. Finally, also $\phi$ can be constructed,
similarly to $\psi$. 
We can introduce an Exotic $E2$-brane intersecting with ordinary ones.
In this way, we generate interactions between Grassmann moduli (or modulini),
living between $E2-D6$ intersections, and ordinary superfields. 
Let us discuss the consistency of the hypercharges
in a construction like the one suggested in Fig.7. 
For intersecting  D-brane model considered in Fig.7,
$U(1)_{Y}$ is defined as a linear 
of $U(1)$ stacks:
\be{Uhyper}
U(1)_{Y}=c_{1}U(1)_{1}+c_{1}'U'(1)+c_{3}U(1)_{3}
\ee
where $U(1)_{2}\subset U(2)$, $U(1)_{2} \subset U(2)$.
So the hypercharge is  
a combination of four abelian charges.
From (\ref{Uhyper}), a consistent assignation of hypercharges,
$Y(\mathcal{X})=-Y(\mathcal{Y})=-2/3$, $Y(\Psi)=0$,
and the ones of
SM particles, can be found. 
In particular, we find 
$c_{3}=1/3$, $c_{1}=c_{1}'=-1$.
Such a combination could not satisfy 
consistency with cycles, but 
at least we can extend our 
model with other extra $U(1)$s
for consistency of hypercharge combination.
Also the presence of possible flavor branes can 
change such an estimation. 
Of course, the general idea remains 
valid. 
As in \cite{Addazi:2014ila}, a non-perturbative mass term 
between ${\b X},{\bf Y}$ can be generated by two mixed-disk amplitudes,
shown in Fig.7-(b).
In fact, from these, 
\be{L}
\mathcal{L}_{E2-D6-D6''}\sim \nu \tau^{i} {\bf X}_{i}+{\bf Y}_{ij}\tau^{i}\tau^{j}
\ee
where $i,j$ are the color indices of the $U(3)$-stack. 
A new superpotential term, not allowed at perturbative level, 
is obtained, 
integrating out modulini:
\be{WE2}
\mathcal{W}_{E2}=M_{S}e^{-S_{E2}}\int d^{3}\tau d\omega e^{\nu \tau^{i} {\bf X}_{i}+{\bf Y}_{ij}\tau^{i}\tau^{j}}=M_{S}e^{-S_{E2}}\epsilon_{ijk}{\bf X}^{i}{\bf Y}^{jk}
\ee
where $M_{S}$ is the String scale and $e^{-S_{E2}}$ is the parameterize by geometric moduli of the 3-cycles wrapped by the Euclidean D2-brane in the Calabi-Yau $CY_{3}$. 
As shown Fig.8, an exotic mass term can be generated, in a supersymmetric model, as a loop of susy partners $\psi_{\mathcal{X}},\psi_{\mathcal{Y}}$ and a gaugino (gluino, zino or photino), 
with $\mathcal{M}_{0}^{2} \sim m_{\tilde{g}}M_{S}e^{-S_{E2}}$, $m_{\tilde{g}}$ gaugino mass. 

We would like to note that all contributions on irreducible gauge anomalies,  cancel each other, in this D-brane construction.
In fact, $\mathcal{X},\mathcal{Y}$ do not 
introduce extra anomalous contributions 
with respect to SM fields content.
For instance, $SU(3)^{3}$ 
anomalies give equal and opposite contributions
because of
$Tr[\mathcal{X}]=1$ and $Tr[\mathcal{Y}]=N_{c}-4=-1$.
On the other hand, anomalous extra $U(1)$ are introduced with respect to SM gauge group:
new $Z'$ are introduced as in any string-inspired model, with masses generated 
by a St$\ddot{u}$ckelberg mechanism \cite{212,215}. 
Anomalies that could appear as a serious problem in gauge models, 
are cancelled by Generalized Chern-Simons (GCS) terms 
as a generalized Green-Schwarz mechanism \cite{216, 217}
\footnote{The St\"uckelberg mechanism has a lot of different 
intriguing applications. Let us mention that
 a Lorentz Violating
 Massive gravity can be realized through a St\"uckelberg mechanism \cite{LIV1,LIV2,LIV3}. Recently, geodetic instabilities of St\"uckelberg
 Lorentz Violating
 Massive gravity were discussed in \cite{Addazi:2014mga}. }

Finally, we would like to remark that, an exotic mass term (\ref{WE2}) 
cannot be introduced by-hand, at perturbative level, because of R-parity, 
i.e R-parity is dynamically broken, without the generation 
of other dangerous R-parity violating operators, as explained in \cite{Addazi:2014ila,Addazi:2015rwa,Addazi:2015hka}.

\section{Conclusions}

In this paper, we have discussed a simple alternative model generating
a Majorana mass for the neutron, connecting Majorana's proposal 
to deep issues regarding Baryogenesis and Dark Matter. 
In particular, we have introduced just one exotic vector-like pair of color-triplet scalars,
a sterile Majorana fermion $\psi$, and a scalar giving mass to $\psi$. 
An exotic vector-like pair is characterized by an extra peculiar mass term, violating baryon number by $\Delta B=1$.
In particular, we got limits on exotic mixing mass 
parameter from LHC physics. 
We have seen how Baryogenesis can be realized, also during the post-sphaleron epoch,
and we predict a neutron-antineutron transition with a time
interesting for the next generation of experiments: $\tau_{n\bar{n}}\sim 300\, \rm yr$.
We have also considered, an alternative scenario, in which the sterile fermion is a metastable WIMP-like particle. 
In this case, a neutron-antineutron transition can be generated by two $\Delta B=1$ oscillations, $n-\psi$ and $\psi-\bar{n}$. 
Finally, we have also shown a possible completion and explanation of such a toy-model,
in which the exotic mass term is generated by non-perturbative exotic stringy istantons.  

We conclude that this model, postulating an exotic vector-like pair of color-triplet scalars, 
deserves attention for its peculiarity and simplicity,
especially considering its possible connections with fundamental issues and its 
implications in B-violations phenomenology such as neutron-antineutron physics and LHC.

\vspace{0.5cm} 

{\large \bf Acknowledgments} 
\vspace{3mm}

A.A would like to thanks Galileo Galilei Institute for Theoretical Physics 
for the hospitality, where this paper was prepared. 
I would like to thank Massimo Bianchi, Zurab Berezhiani and Luca Di Luzio for interesting conversations. I also would like to thank the anonymous 
referee for his important comments and suggestions.
The work of A.A. was supported in part by the MIUR research
grant "Theoretical Astroparticle Physics" PRIN 2012CPPYP7.

% *** Biliography ***

\end{document}